\documentclass [12pt]{article}

\begin{document}

\vspace{1cm}

\begin{center}

{\large\bf Relation of Atmosphere Surface Pressure Fluctuations
with Geomagnetic and Cosmic Factors}

\vspace{1cm}

{\it M.B.Bogdanov, A.V.Fedorenko}

\vspace{1cm}

{\footnotesize 
Saratov State University, ul.Astrakhanskaya 83, Saratov, 410026 Russia}

{\footnotesize 
e-mail: BogdanovMB@info.sgu.ru; FedorenkoAV@info.sgu.ru }

\end{center}

\vspace{1cm}

We present results of the digital spectral analysis of the
time series con\-tain\-ing the daily samples of the atmosphere
surface pressure measured at Saratov city from Jan 1, 1995
to Dec 31, 1999. We calculated also cross correlation
functions between the pressure fluctuations and the
planetary geomagnetic indices, $A_p$ and $C_p$, and cross
correlation function between these fluctuations and the
cosmic-ray intensity. Two peaks detected in the cal\-cu\-la\-ted
power spectrum are related, probably, with the atmosphere
tidal waves. The relation of the pressure fluctuations
with the $A_p$ index of the geomagnetic activity is
statistically insignificant. The cross correlation func\-tion
between the pressure fluctuations and the $C_p$ index has
the maximal value $0.044 \pm 0.023$ at the pressure time
delay $4^d$. The statistically significant negative
correlation is discovered between the surface pressure
fluctuations and the cosmic-ray intensity. The cross
correlation function has two minima: at the zero time delay
with the value $-0.068 \pm 0.023$, and at the pressure
time delay $12^d$ with the value $-0.087 \pm 0.023$. The
negative correlation between the pressure and the
cosmic-ray intensity is observed at the pressure time delay
as long as $17^d$ and can be explained by the so-called
condensation mechanism of an interaction of the cosmic rays
with the atmosphere.

\vspace{0.5cm}

\begin{center}

{\bf   1. Introduction }

\end{center}

\vspace{0.5cm}

The problem of an investigation of the influence of cosmic
factors on the Earth's atmosphere attracts the attention of
many researchers. Among these factors there are tidal
interactions with the Moon and the Sun as well as the
nutational motion of the Earth's rotation axis and the
proper nutation of the Earth (motion of the pole) [1].
Cosmic rays of solar and galactic origin also interact with
the atmosphere. The fluxes of these cosmic rays depend on a
state of the Earth's magnetosphere and, as a consequence,
on the solar activity [2]. The aim of our current study is
the investigation of possible relation of the atmosphere
surface pressure fluctuations with geomagnetic and cosmic
factors using the digital spectral analysis of time series
and the cross correlation functions calculations.

\vspace{0.5cm}

\begin{center}

{\bf 2. Observational data }

\end{center}

\vspace{0.5cm}

We examined the time series containing the daily samples of
the atmos\-phe\-re surface pressure measured by the mercury
barometer at Saratov city ($51^\circ.5$ N; $46^\circ.0$ E;
$h = 80$ m) from Jan 1, 1995 to Dec 31, 1999 at 12 hours of
third time zone. As a first stage, we subtracted from the
pressure samples the main annual harmonic with the period
equal to the tropical year ($365.^d2422$). The amplitude $A
= 5.36$ hPa and the phase $\phi = 1.44$ rad. of this
harmonic have been determined by fitting to the time series
a sinusoid using the least squares method.

Cosmic rays are not registered at Saratov city. However the
flux of galactic cosmic rays of high energies which can to
penetrate in atmosphere down to the sea level has a high
correlation at different points of some region. Therefore we
used the daily data on cosmic rays measured at Apatity
station. The cosmic-ray flux samples have been filtered by
the moving average digital filter using the averaging
interval time length $200^d$. 

The data on $A_p$ and $C_p$ planetary geomagnetic indices accepted
from the Space Physics Interactive Data Resource
(http://spidr.ngdc.noaa.gov/).

\vspace{0.5cm}

\begin{center}

{\bf 3. Calculation and analysis of power spectrum}

\end{center}

\vspace{0.5cm}

We used for the spectral analysis of the time series
containing the daily samples of atmosphere surface pressure
with subtracted main annual harmo\-nic the classical version
of the power spectrum estimation based on calcu\-la\-tion of the
Fourier transform of an autocorrelation function with the
Tukey's spectral window [3]. The maximal time delay of the
autocorrelation function was chosen $450^d$. In this case
the number of degrees of freedom equal to $11$, and the
half-power bandwidth is $0.00296$ cycles per day. The upper
frequency boundary of the calculated spectrum was chosen
$0.10$ cycles per day.

Several well-defined peaks there are in the power spectrum.
We adopted for periods of feasible harmonics which
correspond to these peaks to be equal to the periods of peak
maximal points and the error of the estimation of a period
to be determined by a value of the half-power bandwidth. Two
peaks with the periods $29^d.9 \pm 2^d.7$ and $16^d.4 \pm
0^d.8$, probably, are the tidal harmonics $M_m (27^d.55)$
and $M_f (13^d.66)$, respectively [1]. The peaks with
periods $45^d.4 \pm 6^d.1$ and $11^d.8 \pm 0^d.4$, probably,
are related with the modes of proper atmosphere
oscillations [4].

\vspace{0.5cm}

\begin{center}

{\bf 4. Investigation of correlation between surface pressure} 

{\bf and geomagnetic activity}

\end{center}

\vspace{0.5cm}

We calculated cross correlation functions between the
pressure fluc\-tu\-ati\-ons and the planetary geomagnetic indices
$A_p$ and $C_p$ for different time delays of the pressure measured
in integer days numbers. The relation of the pressure
fluctuations with the $A_p$ index of geomagnetic activity is
statistically in\-sig\-ni\-fi\-cant. The cross correlation function
between the pressure fluctuations and the $C_p$ index has a
maximum value $0.044 \pm 0.023$ at the pressure time delay
$4^d$. This result is in a good accordance with the data of
Mustel et al. [5].

It is clear that the magnetic field of the Earth not affect
directly on the atmosphere air. A relation of atmosphere
parameters variations with the geomagnetic activity is a
consequence of its relation with factors of the solar
activity including the X-ray and the UV-radiation of solar
flares, the solar cosmic rays, and the modulation of a
galactic cosmic-ray flux. Therefore it is interest to
examine a relation of the pressure fluctuations with the
primary factors of the solar activity.

\vspace{0.5cm}

\begin{center}

{\bf 5. Investigation of correlation between surface pressure}

{\bf and cosmic-ray intensity}

\end{center}

\vspace{0.5cm}

According to the modern conception, the main canal of the
solar activity influence on the Earth's troposphere is a
modulation of the flux of galactic cosmic rays of high
energies which can to penetrate in atmosphere down to the
sea level [2]. Investigations of correlations between the
atmosphere characteristics and the cosmic-ray intensity can
potentially elucidate the mechanism of an interaction of
cosmic rays with the Earth's atmosphere. 

We calculated the cross correlation function between the
pressure fluctu\-ati\-ons and the cosmic-ray intensity. The
values of this function are negative and small enough, but
mostly are statistically significant. The cross cor\-re\-la\-tion
function has two minima: at the zero time delay (apparently,
a moment of the Forbush decreases in cosmic-ray intensity)
with the value $-0.068 \pm 0.023$, and at the pressure time
delay $12^d$ with the value $-0.087 \pm 0.023$. The
negative correlation between the pressure and the cosmic-ray
intensity is observed at the pressure time delay as long as
$17^d$. This result is in a good accordance with the data of
Grigor'ev et al. [6], who studied the variation of surface
pressure after the Forbush decreases of cosmic-ray intensity
at Yakutsk city using the epoch superposition method. These
authors have found that the response of the pressure
variations to the Forbush decreases has on average duration
$13^d$ and amplitude $+2$ hPa, respectively.

The negative correlation between the surface pressure and
the cosmic-ray intensity can be explained by the so-called
condensation mechanism of an interaction of cosmic rays with
the atmosphere air. Cosmic rays can produce hygroscopic
clusters in the lower atmosphere through ionization. This
must result in a certain increase and decrease in pressure
if the cosmic-ray intensity decreases and increases,
respectively.

It is interesting that negative correlation between the
surface pressure and the cosmic-ray intensity exists also
for negative pressure time delays down to $-5^d$. This
effect can be connected partly with the finite duration of
the Forbush decrease. Another explanation of the forward
negative correlation is an influence of a microwave
radiation of solar flares. This radiation also can produce
the water molecular clusters in the lower layers of the
atmosphere [7].

\vspace{0.5cm}

\begin{center}

{\bf References}

\end{center}

\vspace{0.5cm}

{\parindent=0pt

1. Maksimov I.V. Geophysical forces and ocean waters.
Gydrometeoizdat, Leningrad, 1970 (in Russian).

2. Avdyushin S.I.,Danilov A.D. Geomagnetism and Aeronomy.
2000. V.40. P.545.

3. Jenkins G.M., Watts D.G. Spectral analysis and its
applications. Holden-Day, Inc., San Francisco, 1968.

4. Tudrij V.D.,Kolobov N.V. Fluctuations of cyclonic
processes in northern hemisphere of the Earth.
Gydrometeoizdat, Leningrad, 1984 (in Russian).

5. Mustel E.R,Chertoprud V.E.,Mouloukova N.B. Scientific
Information of Astronomical Council of the USSR Academy of
Sciences. 1986. V.60. P.161 (in Russian).

6. Grigor'ev V.G.,Bezrodnykh I.P.,Morozova E.I.,
Timofeev V.E.,Skryabin N.G. Geomagnetism and Aeronomy. 2002.
V.42. P.520.

7. Kondrat'ev K.Ya., Nikol'skii G.A. Investigation of the
Earth from Space. 1995. No 5. P.3 (in Russian).

}

\end{document}